\documentclass[aps,prb,twocolumn,floatfix]{revtex4}

\usepackage{makecell}
\usepackage{graphicx}
\usepackage{dcolumn}
\usepackage{bm}
\usepackage{times}
\usepackage{color}
\usepackage{appendix}

\begin{document}
\bibliographystyle{apsrev4-1}
\title{Planar Hall effect in the Dirac semimetal PdTe$_2$}

\author{Sheng Xu}\thanks{These authors contributed equally to this paper}
\author{Huan Wang}\thanks{These authors contributed equally to this paper}
\author{Xiao-Yan Wang}
\author{Yuan Su}
\author{Peng Cheng}
\author{Tian-Long Xia}\email{tlxia@ruc.edu.cn}

\affiliation{Department of Physics, Renmin University of China,
Beijing 100872, P. R. China} \affiliation{Beijing Key Laboratory of
Opto-electronic Functional Materials $\&$ Micro-nano Devices, Renmin
University of China, Beijing 100872, P. R. China}

\date{\today}

\begin{abstract}
We report the synthesis and magneto-transport measurements on the
single crystal of Dirac semimetal PdTe$_2$. The de Haas-van Alphen
oscillations with multiple frequencies have been clearly observed,
from which the small effective masses and nontrivial Berry phase are
extracted, implying the possible existence of the Dirac fermions in
PdTe$_2$. The planar Hall effect and anisotropic longitudinal
resistivity originating from the chiral anomaly and nontrivial Berry
phase are observed, providing strong evidence for the nontrivial
properties in PdTe$_2$. With the increase of temperature up to 150
K, planar Hall effect still remains. The possible origin of mismatch
between experimental results and theoretical predictions is also
discussed.

\end{abstract}
\maketitle

\section{Introduction}

Topological Dirac/Weyl semimetals have attracted extensive attention
in condensed matter physics community because of their novel
properties\cite{liu2014discovery,xiong2015evidence,neupane2014observationCd3As2,YLChen2014stableCd3As2,PhysRevLett.113.027603,liang2015ultrahigh,li2015giant,li2015negative,PhysRevX.5.011029}.
In the Dirac semimetals, the fourfold degenerate band crossings in
the vicinity of the Fermi level are known as Dirac
points\cite{wehling2014dirac}. As the inversion symmetry or time
reversal symmetry breaks, Dirac point degenerates into Weyl points
with opposite chirality\cite{PhysRevB.83.205101}. Interestingly, the
parallel magnetic and electric field pumps electrons between Weyl
nodes with opposite chirality, which leads to a chiral current that
contributes to the negative magnetoresistance
(NMR)\cite{huang2015observation,zhang2016signatures,son2013chiral}.
The observation of the NMR has been regarded as a routine method to
study Dirac/Weyl fermions in transport measurements. However, the
NMR may also be contributed by other
mechanisms\cite{ritchie2003magnetic,fauque2013two,pippard1989magnetoresistance,hu2005current,kikugawa2016interplanar,osada2008negative,tajima2009effect},
and sometimes it is hard to clearly observe the chiral anomaly
induced NMR in some topological semimetals with the coexisting
positive orbital MR. The nontrivial Berry phase extracted from the
quantum oscillations provides another evidence for the
identification of the topological characteristics. However, it is
difficult to extract the Berry phase from those materials with
complex band structures or the materials without observable quantum
oscillations.

It is worth noting that planar Hall effect (PHE), another
characteristic induced by the chiral anomaly and nontrivial Berry
phase, is predicted as a proof of the existence of Weyl fermions in
the topological semimetals when the applied magnetic and electric
fields are coplanar\cite{burkov2017giant,nandy2017chiral}, which is
 already observed in GdPtBi\cite{kumar2018planar}/DyPdBi\cite{Pavlosiuk2018Planar},
VAl$_3$\cite{singha2018planar},
Mo/WTe$_2$\cite{liang2018origin,chen2018planar,wang2018planar},
Cd$_3$As$_2$\cite{li2018giant,wu2018probing} and
ZrTe$_{5-\delta}$\cite{li2018giant2}, TaP\cite{yang2018giant}. On
the other hand, the PHE with a smaller value is also observed in
some topological trivial ferromagnetic metals, which originates from
the interplay of magnetic order and spin-orbit interaction
\cite{nazmul2008planar}. Thus, PHE is believed as a direct evidence
in magneto-transport studies to identify the nontrivial topological
property in nonmagnetic materials, especially for those candidates
in which band structures are complex, oscillations are absent or the
NMR is difficult to observe.

The transitional metal dichalcogenide PdTe$_2$ is believed to be
type-II Dirac semimetal as
PtTe$_{2}$\cite{wang2016hass,Noh2017Experimental,fei2017nontrivial,zheng2018detailed,yan2017lorentz}.
In this paper, we report the magneto-transport measurements of
nonmagnetic PdTe$_2$ with the de Haas-van Alphen (dHvA) oscillations
observed, from the analysis of which the multiple frequencies,
effective masses and nontrivial Berry phase are extracted. The PHE
and anisotropic longitudinal resistivity are observed, which
provides a strong evidence for the existence of topological
nontrivial properties in PdTe$_2$.

The single crystals of PdTe$_2$ were grown by melting the mixture of
Pd and Te powder with a ratio 1:2.2 in a sealed evacuated quartz
tube at 800$^\circ \mathrm{C}$ for 2 days, then slowly cooling down
in 7 days to 500$^\circ \mathrm{C}$, and remaining for 7 days before
turning off the furnace. The pattern of XRD was collected from a
Bruker D8 Advance x-ray diffractometer using Cu K$_{\alpha}$
radiation. The magneto-transport measurements were performed on a
Quantum Design physical property measurement system (QD PPMS).

\section{Results and discussion}

\begin{figure*}[htbp]
\centering
\includegraphics[width=0.8\textwidth]{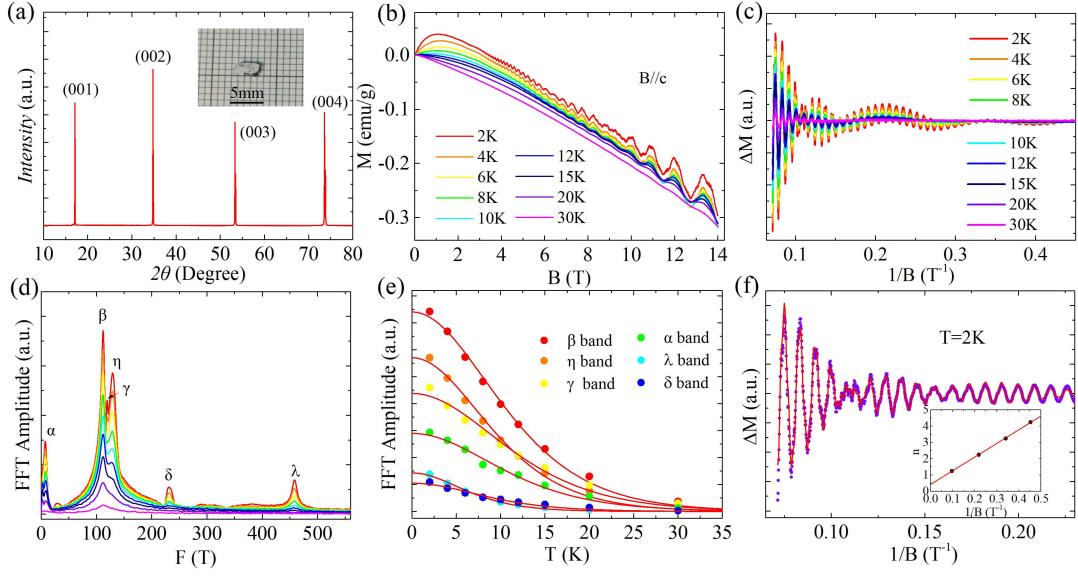}
\caption{(a) XRD of the PdTe$_2$ single crystal. Inset shows the
picture of a grown crystal. (b) The dHvA oscillations of PdTe$_2$ at
various temperatures. (c) The amplitude of dHvA oscillations plotted
as a function of 1/B. (d) FFT spectra of the oscillations between 5
K and 30 K. (e) The temperature dependence of relative FFT
amplitudes of each frequency and the fitting results by $R_T$. (f)
The LK formula fitting of the dHvA oscillations excluding the
$\alpha$ band. The inset shows the LL index fan diagram for $\alpha$
band.}
\end{figure*}

The x-ray diffraction (XRD) pattern of single crystal with strongly
(\emph{00l}) peaks is shown in Fig.1 (a), which indicates that the
surface of the crystal is the ab plane. The inset of Fig.1 (a) shows
a typical picture of grown PdTe$_2$ crystal with metallic luster.
Clear dHvA oscillations are observed in PdTe$_2$ crystals at the
temperature range 2 K - 30 K with the magnetic field parallel to the
c-axis (B//c) as shown in Fig.1 (b). With the temperature
increasing, the oscillations gradually weakens. After subtracting a
smoothing background, the oscillatory amplitudes of magnetization
against 1/B were plotted in Fig.1 (c). The extracted frequencies
from the fast Fourier transform (FFT) analysis are shown in Fig.1
(d). The oscillations can be well described by the Lifshitz-Kosevich
(LK) formula\cite{shoenberg2009magnetic},

\begin{equation}\label{equ1}
\centering
\Delta M \propto -B^{1/2}\frac{\lambda T}{sinh(\lambda T)}e^{-\lambda T_D}sin[2\pi(\frac{F}{B}-\frac{1}{2}+\beta+\delta)]
\end{equation}

\noindent where $\lambda=(2\pi^2k_{B}m^*)/(\hbar eB)$, $T_D$ is the
Dingle temperature and $\beta=\Phi_B /2\pi$ ($\Phi_B$ is the Berry
phase). The phase shift $\delta$ is determined by the
dimensionality, $\delta=0$ and $\delta= \pm1/8$ for 2D and 3D
system, respectively. The thermal factor $R_T=(\lambda
T)/sinh(\lambda T)$ in LK formula is employed to fit the temperature
dependence of the oscillatory amplitude (Fig.1 (e)). The effective
masses around 0.04-0.08$m_e$ obtained from the fitting are listed in
Table I. The Berry phase $\Phi_B=2\pi \beta$ can be extracted from
the fitting of the LK formula or the analysis of Landau level (LL)
index fan diagram. We adopt the latter method to extract the Berry
phase of $\alpha$ band, and the corresponding LL index fan diagram
is shown in the inset of Fig.1 (f).  The Landau index of the dHvA
oscillations maximum should be n$+$1/4\cite{hu2016evidence}, and the
LL index fan diagram gives an intercept of 0.43 which is consistent
with previous reports\cite{zheng2018detailed,fei2017nontrivial}.
Thus, the Berry phase of $\alpha$ band is obtained to be 2.11$\pi$
for $\delta=-1/8$ or 1.61$\pi$ for $\delta=1/8$, respectively. Both
of them are close to 2$\pi$, which indicates the topological trivial
character of $\alpha$ band. After filtering the frequency of
$\alpha$ band, we obtained the high frequency oscillations as shown
in Fig.1 (f) (violet dots). The Berry phases are extracted from the
fitting of the multiband LK formula (red line), and the
corresponding values are listed in Table I. Several of the Berry
phases are close to the nontrivial value $\pi$ indicating the
possible existence of Dirac fermions in PdTe$_2$.

\begin{figure}[htbp]
\centering
\includegraphics[width=0.48\textwidth]{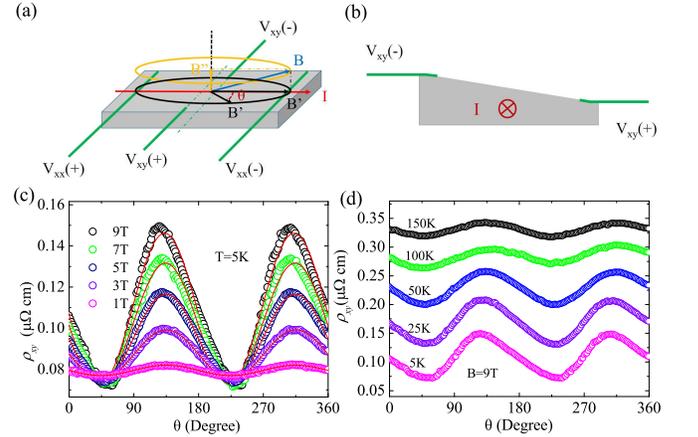}
\caption{(a) Schematic diagram for the planar Hall resistivity
measurements and the misalignments geometry. (b) Lateral view of the
schematic diagram in (a). (c) The angular dependence of $\rho_{xy}$
under various magnetic fields at 5 K. (d) The planar Hall
resistivity $\rho_{xy}$ as a function of angle at different
temperatures (B=9 T).}
\end{figure}

\begin{figure}[htbp]
\centering
\includegraphics[width=0.48\textwidth]{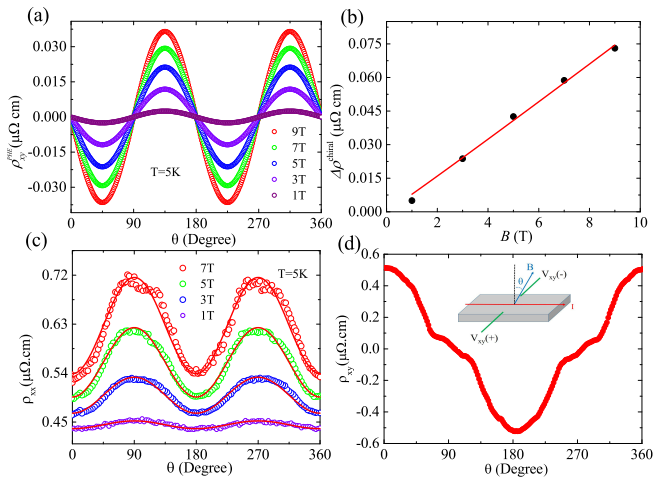}
\caption{(a) The extracted intrinsic planar Hall resistivity
$\rho^{PHE}_{xy}$ versus angle $\theta$ under different magnetic
fields (T=5 K). (b) The extracted parameter $a$ varies linearly with
magnetic field at T=5 K. (c) The anisotropic longitudinal
resistivity under different magnetic fields at 5 K. (d) The angular
dependence of the normal Hall resistivity at 10 K. The inset shows
the schematic diagram for the normal Hall resistivity measurement.}
\end{figure}

The PHE is also examined to further study whether the topological
nontrivial states exist in PdTe$_2$. The planar Hall resistivity
($\rho^{PHE}_{xy}$) and anisotropic longitudinal resistivity
($\rho_{xx}$) measurements with the coplanar magnetic and electric
field are reported in details. As the theory predicts, the angular
dependence of PHE and anisotropic longitudinal resistivity in the
Dirac/Weyl semimetals can be described as,

\begin{equation}\label{equ2}
\centering
\rho^{PHE}_{xy}= -\Delta \rho^{chiral} sin\theta cos\theta
\end{equation}
\begin{equation}\label{equ3}
\centering
\rho_{xx}= \rho_{\perp}-\Delta \rho^{chiral} cos^2{\theta}
\end{equation}
where $\rho^{PHE}_{xy}$ is the planar Hall resistivity, and $\Delta
\rho^{chiral}=\rho_{\perp}-\rho_{\parallel}$ is the chiral anomaly
induced resistivity component when the current and the magnetic
field are coplanar ($\rho_{\perp}$ or $\rho_{\parallel}$ represents
the resistivity with the magnetic field perpendicular or parallel to
the current, respectively). $\theta$ is defined as the angle between
the current and magnetic field direction. $\rho_{xx}$ corresponds to
the angle-dependent longitudinal resistivity. The results can not be
well described with the Eqs. (2) and (3). Considering the actual
difficulties in the measurements, the disagreement is attributed to
three types of misalignments. The first one is that the current and
the magnetic field B(cyan line in Fig.2 (a)) may not be completely
coplanar due to the tilt of the sample. Hence, an additional
magnetic component B''(yellow line in Fig.2 (a)) normal to the
sample surface is induced and leads to an unexpected normal Hall
resistance. The parallel component B' rotates in the sample plane
with the angle $\theta$ from the direction of the current, as
illustrated in Fig.2 (a). It is fortunate that the normal Hall
resistivity complies with an odd function versus magnetic field
while planar Hall resistivity is not (even function). The influence
of the normal Hall resistance can be easily subtracted by taking an
average of the data obtained under the positive and negative
magnetic field. The second type of misalignment originates from the
asymmetrical Hall contacts which introduce an extra longitudinal
resistivity and can be described as $b cos^2{\theta}$. The last one
is attributed to the non-uniform thickness of the sample, which
leads to a constant Hall resistivity. To demonstrate the analysis
clearly, a schematic diagram is presented to show the planar Hall
resistivity measurement and the misalignments as displayed in Fig.2
(a). The lateral view exhibiting non-uniform thickness of the sample
is also shown in Fig.2 (b). Thus, Eq. (2) is modified as,

\begin{equation}\label{equ4}
\centering
\rho_{xy}= -a sin\theta cos\theta+ b cos^2{\theta}+ c
\end{equation}
\noindent the first part is the intrinsic PHE term $\rho^{PHE}_{xy}$
induced by the chiral anomaly and the nontrivial Berry phase. The
second and third are modified terms corresponding to the latter two
kinds of misalignments.

Figure 2(c) shows the angular dependence of $\rho_{xy}$ under
various magnetic field at 5 K, which have eliminated the extra
normal Hall resistivity by taking the average of the values obtained
under positive and negative fields. The observed $\rho_{xy}$
exhibits the valley at $\pi/4$ and the peak at $3\pi/4$ with the
period of $\pi$, which is well fitted by the Eq. (4). The fitting
results are demonstrated as the red curves in Fig.2 (c). Even though
the amplitude of the PHE decreases with the increasing temperature,
the PHE is still observed up to 150 K.

\begin{table}
\centering \caption{Parameters derived from the dHvA oscillations.
$F$, oscillation frequency; $m^*$, effective mass; $\Phi_B=2\pi
\beta$, Berry phase.} \label{oscilltion}
\begin{tabular}{cccccc}
   \hline\hline
     & $F$ (T) & $m^*/m_e$ & \makecell[c]{$\Phi_B=2\pi\beta$ \\ ($\delta=1/8$)} & \makecell[c]{$\Phi_B=2\pi\beta$ \\ ($\delta=-1/8$)} & \makecell[c]{$\Phi_B=2\pi\beta$ \\ ($\delta=0$)}\\
   \hline
   $\alpha$ & 7.4 & 0.05 & 2.11$\pi$ & 1.61$\pi$ & / \\
   $\beta$ & 112.0 & 0.05 & 0.01$\pi$ & 1.51$\pi$ & 1.76$\pi$ \\
   $\gamma$ & 119.3 & 0.04 & 1.11$\pi$ & 0.61$\pi$ & 0.86$\pi$ \\
   $\eta$ & 129.3 & 0.05 & 1.30$\pi$ & 0.80$\pi$ & 1.05$\pi$ \\
   $\delta$ & 230.9 & 0.08 & 0.54$\pi$ & 0.04$\pi$ & 0.29$\pi$ \\
   $\lambda$ & 458.6 & 0.06 & 1.84$\pi$ & 1.33$\pi$ & 1.59$\pi$ \\
\hline\hline
\end{tabular}
\end{table}
The intrinsic PHE extracted from the fitting by Eq. (4) is plotted
in Fig.3 (a). The amplitudes of PHE at different field are displayed
in Fig.3 (b), which shows a \textbf{B} dependence. In addition, we
plotted the anisotropic longitudinal resistivity under different
magnetic field at 5 K as shown in Fig.3 (c). The observed
$\rho_{xx}$ demonstrates the period of $\pi$ with maximum at $\pi/2$
and $3\pi/2$ when the magnetic field is applied normal to the
current. The curves in red shown in Fig.3 (c) represent the fitting
by Eq. (3), which coincide well with the experimental results. To
clarify the difference from the PHE, the normal Hall measurement is
applied. Fig.3 (d) shows the angular dependence of the normal Hall
resistivity at 10 K demonstrating the period of $2\pi$, twice as
much as that in PHE.

\section{Summary}
In conclusion, we have grown high quality single crystals of
PdTe$_2$ and investigated the magneto-transport properties. The dHvA
oscillations with multiple frequencies have been observed. According
to the analysis of the oscillations, we obtained small effective
masses and nontrivial Berry phase, which suggest the possible
existence of Dirac/Weyl fermions. Also, a clear signal of PHE and
the anisotropic longitudinal resistivity are observed, which can be
regarded as the result of the chiral anomaly and nontrivial Berry
phase. Besides, taking the noncoplanar magnetic field with the
sample, non-uniform sample thickness and asymmetrical Hall contacts
into consideration, the modified model agrees well with the data
obtained. Thus, the nontrivial Berry phase extracted from the dHvA
oscillations indicates the topological nontrivial characteristic of
PdTe$_2$, while the PHE provides the further evidence.

\section{Acknowledgments}
This work is supported by the National Natural Science Foundation of
China (No.11574391, No.11874422), the Fundamental Research Funds for
the Central Universities, and the Research Funds of Renmin
University of China (No.18XNLG14).
\bibliography{Bibtex}
\end{document}